\documentclass[11pt]{article}

\usepackage{psfig}
\usepackage{amssymb}
\usepackage{amsmath}

\begin{document}
\title{\bf An Invariant Charge Model for All
$\mathbf{q^2>0}$ in QCD and  Gluon Condensate}

\author{A.I. Alekseev\thanks{E-mail: alekseev@mx.ihep.su} \\
{\small Institute for High Energy Physics, 142281 Protvino, Moscow
Region, Russia}\\ B.A. Arbuzov\thanks{E-mail: arbuzov@theory.sinp.msu.ru} \\
\small Skobeltsyn Institute for Nuclear Physics, Moscow State
University\\ \small 119899 Moscow, Russia}

\date{}

\maketitle

\thispagestyle{empty}

\begin{abstract}
Under assumption of singular behavior of invariant charge
$\alpha_s(q^2)$ at $q^2\simeq 0$ and of large $q^2$ behavior,
corresponding to the perturbation theory up to four loops, a
procedure is considered of smooth matching the $\beta$-function
at a boundary of perturbative and nonperturbative regions. The
procedure  results in a model for $\alpha_s$ for all $q^2>0$ with
dimensionless  parameters being fixed and  dimensional parameters
being expressed in terms of only one quantity $\Lambda_{QCD}$.
The gluon condensate which is defined by the nonperturbative part
of the invariant charge is calculated for two variants of ``true
perturbative" invariant charge, corresponding to freezing option
and to analytic one in nonperturbative region. Dimensional
parameters are fixed  by varying normalization condition
$\alpha_s(m^2_{\tau}) = 0.29,\,0.30,\,...,\, 0.36$. It is
obtained that on the boundary of perturbative region $\alpha_0=$
$\alpha_s(q_0^2)\simeq$ 0.44, the procedure results in
nonperturbative Coulomb component
$\alpha_{\mathrm{Coulomb}}\simeq$ 0.25, the nonperturbative region
scale $q_0\simeq$ 1 GeV, the model parameter $\sigma\simeq$ (0.42
GeV)$^2$ which  suits  as string tension parameter, the gluon
condensate appears to be  close  for two variants considered,
$K\simeq$ (0.33 -- 0.36 GeV)$^4$ (for $\alpha_s(m^2_{\tau}) =
0.33$).\\


\end{abstract}
\vfill \eject

\section{Introduction}

A consistent description of interaction of fundamental fields both
at large and short distances is one of the most important
problems of QCD. The strength of the interaction is defined by
invariant charge $\alpha_s(q^2)$ (the running coupling constant),
which satisfies the renormalization group equation. The purpose
of the present work consists in a formulation of a model for
$\alpha_s(q^2)$ for all $q^2 > 0$, which is appropriate for
description both of perturbative and of nonperturbative phenomena
and needs minimal number of parameters.\footnote{This paper is an
extended version of Ref.~\cite{0407056}.}

We assume that there exists some value $q_0$, which characterizes
the nonperturbative effects scale and the corresponding value
$\alpha_0=$ $\alpha_s(q_0^2)$ such, that for $q^2 > q_0^2$ and
thus $\alpha_s(q^2) < \alpha_0$ the finite loop perturbation
theory is applicable and sufficient, while nonperturbative
effects prevails in  region $q^2 < q_0^2$, which contains also
nonphysical singularities of the perturbation theory and so here
this theory essentially needs an extension  of  definition.
 In general, the invariant charge and the
beta-function as well are depending on  a renormalization
scheme~\cite{BoShir,Stevenson}. For definiteness   while
performing calculations at $q^2 \geq q_0^2$ we  use
$\overline{MS}$ scheme. It is well-known, that the
$\beta$-function in the perturbative QCD is of the form
\begin{equation}
\beta_{\rm pert}(h)=-b_0 h^2-2b_1h^3-
\frac{b_2}{2}h^4-b_3h^5
+O(h^6), \,\,\, h=\frac{\alpha_s(q^2)}{4\pi}.
\label{1}
\end{equation}
For $n_f=$ 3 we have values of coefficients $b_0=$ 9,
$b_1=$ 32,  $b_2\simeq$
1287.67,  $b_3\simeq$ 12090.38
(coefficients $b_0$,
$b_1$ do not depend on renormalization scheme while
values $b_2$, $b_3$  correspond to a choice of
$\overline{MS}$-scheme). Expressions obtained by
solution of Gell-Mann -- Low equation
\begin{equation}
q^2\frac{\partial h(q^2)}{\partial q^2}=\beta(h)
\label{2}
\end{equation}
with the use of~(\ref{1}), are widely used for sufficiently large
momenta transfer, however they can not be applied in the infrared
region.

As a matter of fact the behavior of $\alpha_s$ at small momenta
till now is an open question. Lattice methods and SD equations
give no an ultimate answer. For  behavior of the invariant charge
$\alpha_s(q^2)$ at $q^2\rightarrow 0$ a number of variants are
considered (see, e.g. Ref.~\cite{ShirTMF02}). Now the most
popular variants for $q^2 \rightarrow 0$ behavior  are: $\alpha_s
\rightarrow 0$, $\alpha_s \rightarrow O(1)$, $\alpha_s$ is
strongly enhanced. We consider the last possibility,
in particular the well-known singular infrared asymptotic behavior
\begin{equation}
\alpha_s(q^2)\simeq \frac{g^2}{4\pi}\frac{M^2}{q^2}, \;\;\;
q^2\rightarrow 0,
 \label{3}
\end{equation}
for a review see, e.g. Ref.~\cite{Arb1} and more recent papers
Refs.~\cite{AA}. Such behavior of the invariant charge
$\alpha_V$  in so-called V-scheme   corresponds to a linear
confining quark-antiquark static potential at that $g^2
M^2\,=\,6\pi\sigma$, where $\sigma$ is the string tension.
Results of some works~\cite{lat} on the lattice study of the
three-gluon vertex  demonstrate a necessity of taking into
account of nonperturbative contributions to the running coupling
constant being of the form~(\ref{3}). In the framework of the
continuous QFT additional arguments in favor of
behavior~(\ref{3}) are also presented in paper~\cite{Gogohia}.
Highly similar singular infrared behavior has recently developed
model for the QCD analytic invariant charge~\cite{NesterJMP03}.

Asymptotic behavior~(\ref{3}) occurs provided
$\beta(h)\rightarrow-h$ for $h\rightarrow \infty$. We consider a
possibility of behavior~(\ref{3})  and assume the following form
of the infrared $\beta$-function
\begin{equation}
\beta(h)=-h+z, \;\;\; h>h_0,
\label{4}
\end{equation}
where $z$ is a constant and $h_0$ corresponds to the boundary
between perturbative and  nonperturbative regions. For $h<h_0$ we
shall use $\beta$-function~(\ref{1}) with a finite number of loops
taken into account. Our recipe for construction of
$\beta$-function for all $h>0$ consists in a smooth matching of
expressions~(\ref{4}) and (\ref{1}) at point $h=h_0$ in
approximations of the perturbation theory up to four loops. The
demand of the $\beta$-function and its derivative to be
continuous uniquely fix free parameters $z$ and $h_0$ of the
global $\beta$-function (the matched one). Note, that the
presence of parameter $z$ in Eq.~(\ref{4}) which corresponds to
Coulomb contribution in invariant charge gives a possibility of
smooth matching.

We are willing to build the model for the invariant charge, which
precisely coincides the perturbation theory in the perturbative
region $q^2
> q_0^2$, while in the nonperturbative infrared region it
provides simple description of main nonperturbative parameters.
The work is organized in the following way. In Section 2 we
obtain the matched solutions for the cases of 1 --- 4 loops.
Dimensionless   parameters of the model are uniquely  defined and
further the solutions are normalized at the scale of the
$\tau$-lepton mass, that leads to definite  values of dimensional
parameters $\Lambda$, $q_0$, $\sigma$. In Section 3 the gluon
condensate is calculated. In doing this we consider first a
possibility of freezing of the perturbative component of
$\alpha_s$ in the infrared region and secondly a possibility of
analytic behaviour of this component in the infrared region.
Section 4 contains concluding remarks.

\section{Construction of One --- Four-Loop Matched Solutions
for All $\mathbf{q^2>0}$}

For an illustration let us consider the most simple
one-loop case. Conditions of matching give two equations
$$
-b_0h_0^2=-h_0+z,
$$
\begin{equation}
-2b_0h_0=-1. \label{5}
\end{equation}
The solution of set~(\ref{5}) reads
\begin{equation}
h_0=\frac{1}{2b_0}, \;\;\; z=\frac{1}{4b_0}.
\label{6}
\end{equation}
We shall normalize perturbative solution
\begin{equation}
\alpha_s(q^2)=\frac{4\pi}{b_0\ln x}, \;\;\;
x=\frac{q^2}{\Lambda^2_{QCD}}, \;\;\; q^2\ge q_0^2
\label{7}
\end{equation}
by value $4\pi h_0$, that gives
\begin{equation}
x_0=q^2_0/\Lambda^2_{QCD}=e^2,
\label{8}
\end{equation}
where $e = 2.71828...$ . Imposing on $\alpha_s(q^2)$ the
natural condition to be continuous at $q^2=q^2_0$, we may
normalize nonperturbative solution of equation~(\ref{2})
\begin{equation}
\alpha_s(q^2)=4\pi\left(\frac{C}{q^2}+z\right), \;\;\;
q^2\le q_0^2
\label{9}
\end{equation}
by $4\pi h_0$ as well. As a result we obtain
\begin{equation}
C=\frac{3\sigma}{8\pi}=q_0^2(h_0-z), \;\;\; c_0\equiv
C/\Lambda_{QCD}^2=x_0(h_0-z). \label{10}
\end{equation}
Equations~(\ref{10}) are correct for one --- four loops, for
one-loop case $c_0= e^2/4b_0$. For final fixation of the solution
for all $q^2>0$ we need to define $\Lambda_{QCD}$ by normalizing
the solution, say, at point $q^2=m_{\tau}^2$, where $m_{\tau}=$
is the mass of the $\tau$-lepton.

For one-loop case we have following simple formulae for quantities
under consideration
$$
\alpha_0=2\pi/b_0,   \,\,\,
\alpha_{\mathrm{Coulomb}}=\pi/b_0,
$$
\begin{equation}
q_0=e\Lambda_{QCD}, \,\,\, \sigma=\frac{2\pi
e^2}{3b_0}\Lambda^2_{QCD},\,\,\,
 K=\frac{3 e^4}{2\pi^2
b_0}\Lambda^4_{QCD}.\label{11}
\end{equation}
The gluon condensate $K$  for frozen perturbative constituent
which is given here for completeness  will be calculated below.
For $n_f=3$ one has $ \alpha_0=0.698$,\ \ \ \
$\alpha_{\mathrm{Coulomb}}$ $=$ $0.349$,\ \ \ \ $q_0$ $=$
$2.72\Lambda_{QCD}$,\ \ \ \ $\sigma$ $=$ $(1.31\Lambda_{QCD})^2$,\
\ \ \ $K$ $=$\\ $(0.980 \Lambda_{QCD})^4$. For normalization
condition, e.g. $\alpha_s(m^2_{\tau})=$ 0.32 one obtains
$\Lambda_{QCD}=$ 0.201 GeV and we have only qualitative agreement
for quantities considered. As we shall see further on, rather
poor results in the one-loop case are marginal, while results for
many-loop cases turn to be much  more promising.

Let us consider multi-loop cases. Solution $h(q^2)$ of
equation~(\ref{2}) for $L=\ln(q^2/\Lambda^2)\rightarrow \infty$
reads as follows
$$
h(q^2)=\frac{1}{b_0 L}\left\{1-\frac{2b_1}{b_0
^2L}\ln L+\frac{4b_1^2}{b_0^4L^2}
\left[\ln^2 L-\ln L-1+\frac{b_0
b_2}{8b_1^2}\right]\right.
$$
$$
-\frac{8b_1^3}{b_0^6L^3}\left[
\ln^3 L
-\frac{5}{2}\ln^2 L-\left(2-\frac{3b_0b_2}
{8b_1^2}\right)\ln L\right.
$$
\begin{equation}
\left.\left.+\frac{1}{2}-\frac{b_0^2
b_3}{16b_1^3}\right]+O\left(\frac{1}{L^4}\right)\right\}.
\label{15}
\end{equation}
Keeping in the expression terms with powers of logarithms
in denominators up to the first, the second, the third
and the fourth, we fix the 1 --- 4-loop approximations of
the perturbation theory for running coupling constant.
It may be written in the form
$$
\alpha_s(q^2)
=4\pi h(q^2)
=\frac{4\pi}{b_0}a(x),
$$
$$
a(x)=\frac{1}{\ln x}-
b\frac{\ln(\ln x)}{\ln^2x}+b^2\left[\frac{\ln^2(\ln x)}{\ln^3x}-
\frac{\ln(\ln x)}{\ln^3x}+\frac{\kappa}{\ln^3x}\right]
$$
\begin{equation}
-b^3\left[\frac{\ln^3(\ln
x)}{\ln^4x}-\frac{5}{2}\frac{\ln^2(\ln x)}{\ln^4x}+(3\kappa+1)
\frac{\ln(\ln x)}{\ln^4x}+\frac{\bar\kappa}{\ln^4 x}\right].
\label{16}
\end{equation}
Here $x=q^2/\Lambda^2$, and coefficient are defined as
follows
\begin{eqnarray}
b&=&\frac{2b_1}{b^2_0},\nonumber\\
\kappa&=&-1+\frac{b_0 b_2}
{8b^2_1},\nonumber\\
\bar\kappa&=&
\frac{1}{2}-\frac{b_0^2b_3}{16b_1^3}.
\label{17}
\end{eqnarray}
Coefficients  $b$, $\kappa$, $\bar\kappa$ depend on
$n_f$.With $n_f=$ 3 we have $b\simeq$ 0.7901, $\kappa
\simeq$ 0.4147, $\bar\kappa \simeq -1.3679$.
In the case of the two-loop approximation for
perturbative  $\alpha_s$ we have the following set of
equations
$$
b_0 h_0^2+2b_1h_0^3=h_0-z,
$$
$$
2b_0 h_0+6b_1h_0^2=1,
$$
$$
\frac{1}{\ln x_0}-
b\frac{\ln(\ln x_0)}{\ln^2x_0}
=b_0h_0,
$$
\begin{equation}
\frac{c_0}{x_0}+z=h_0.
\label{18}
\end{equation}
The set for three loops reads
$$
b_0 h_0^2+2b_1h_0^3+\frac{b_2}{2}h_0^4=h_0-z,
$$
$$
2b_0 h_0+6b_1h_0^2+2b_2h_0^3=1,
$$
$$
\frac{1}{\ln x_0}- b\frac{\ln(\ln
x_0)}{\ln^2x_0}+b^2\left[\frac{\ln^2(\ln x_0)}{\ln^3x_0}-
\frac{\ln(\ln x_0)}{\ln^3x_0}+\frac{\kappa}{\ln^3x_0}\right]
=b_0h_0,
$$
\begin{equation}
\frac{c_0}{x_0}+z=h_0. \label{19}
\end{equation}
The set for four
loops reads
$$
b_0 h_0^2+2b_1h_0^3+\frac{b_2}{2}h_0^4+b_3h_0^5=h_0-z,
$$
$$
2b_0 h_0+6b_1h_0^2+2b_2h_0^3+5b_3h_0^4=1,
$$
$$
\frac{1}{\ln x_0}-
b\frac{\ln(\ln x_0)}{\ln^2x_0}+b^2\left[\frac{\ln^2(\ln
x_0)}{\ln^3x_0}-
\frac{\ln(\ln x_0)}{\ln^3x_0}+\frac{\kappa}{\ln^3x_0}\right]
$$
$$
-b^3\left[\frac{\ln^3(\ln
x_0)}{\ln^4x_0}-\frac{5}{2}\frac{\ln^2(\ln
x_0)}{\ln^4x_0}+(3\kappa+1)
\frac{\ln(\ln x_0)}{\ln^4x_0}+\frac{\bar\kappa}{\ln^4
x_0}\right]=b_0h_0,
$$
\begin{equation}
\frac{c_0}{x_0}+z=h_0.
\label{20}
\end{equation}
From sets of equations~(\ref{18}) -- (\ref{20}) we find values of
$h_0$, $z$, $x_0$ and $c_0$. They are presented in
Table~\ref{tab1}. Taking into account the existing
data~\cite{Data,Bethke,Piv} we finally fix the momentum dependence
of solutions
%
%
\begin{table}[h]
\caption{ The dimensionless parameters $h_0$, $z$,
$x_0=q_0^2/\Lambda_{QCD}^2$, $c_0=C/\Lambda_{QCD}^2$ on the
number of loops, $n_f=$ 3.}
\begin{center}
{\begin{tabular}{@{}ccccc@{}} \hline \hline
  &1-loop &2-loop&3-loop&4-loop
     \\ \hline \hline
$h_0$
 &   0.0556 &   0.0392 &   \hphantom{0}0.0356 &   \hphantom{0}0.0337
\\
$z$
 &   0.0278 &   0.0215 &   \hphantom{0}0.0203 &   \hphantom{0}0.0197
\\
$x_0$
 &   7.3891 &   7.7763 &  10.2622 &  12.4305
\\
$c_0$
 &   0.2053 &   0.1374 &   \hphantom{0}0.1572 &   \hphantom{0}0.1741
\\ \hline \hline
\end{tabular}}
\end{center} \label{tab1}
\end{table}
for a number of values of the running coupling constant at
$\tau$-lepton mass scale $m_\tau$ with the effective number of
flavors $n_f=3$. The values of $\Lambda_{QCD}$ corresponding to
these normalization conditions are presented in Table~\ref{tab2},
values of boundary momentum $q_0=$ $\sqrt{x_0}\Lambda_{QCD}$ are
presented in Table~\ref{tab3}, the string tension parameter
$\sigma= (8\pi c_0/3)\Lambda_{QCD}^2 $ is given in
Table~\ref{tab5}.
\begin{table}[h]
\caption{ Values of the parameter $\Lambda_{QCD}$ (GeV) on loop
numbers and normalization conditions. Normalization conditions:
$\alpha_s(m_{\tau}^2)=$ 0.29, 0.30, ..., 0.36 with $m_{\tau}=$
1.77703 GeV, $n_f=$ 3.} \begin{center}
{\begin{tabular}{@{}ccccc@{}} \hline \hline
$\alpha_s(m_{\tau}^2)$  &1-loop &2-loop&3-loop&4-loop
     \\ \hline \hline
0.29
 &   0.1600 &   0.3168 &   0.2873 &   0.2837
\\
0.30
 &   0.1734 &   0.3370 &   0.3069 &   0.3026
\\
0.31
 &   0.1869 &   0.3568 &   0.3263 &   0.3212
\\
0.32
 &   0.2005 &   0.3762 &   0.3454 &   0.3394
\\
0.33
 &   0.2143 &   0.3951 &   0.3642 &   0.3573
\\
0.34
 &   0.2280 &   0.4136 &   0.3827 &   0.3749
\\
0.35
 &   0.2418 &   0.4315 &   0.4007 &   0.3920
\\
0.36
 &   0.2556 &   0.4490 &   0.4184 &   0.4087
\\  \hline \hline
\end{tabular}}
\end{center}\label{tab2}
\end{table}
\begin{table}[h]
\caption{ Values of the parameter $q_0$ (GeV)  on loop numbers
and normalization conditions. Normalization conditions are the
same as in Table~\ref{tab2}.} \begin{center}
{\begin{tabular}{@{}ccccc@{}} \hline \hline
$\alpha_s(m_{\tau}^2)$  &1-loop &2-loop&3-loop&4-loop
     \\ \hline \hline
0.29
 &   0.4350 &   0.8833 &   0.9203 &   1.0002
\\
0.30
 &   0.4713 &   0.9397 &   0.9831 &   1.0667
\\
0.31
 &   0.5081 &   0.9949 &   1.0452 &   1.1323
\\
0.32
 &   0.5451 &   1.0490 &   1.1065 &   1.1967
\\
0.33
 &   0.5824 &   1.1018 &   1.1667 &   1.2598
\\
0.34
 &   0.6198 &   1.1533 &   1.2258 &   1.3216
\\
0.35
 &   0.6572 &   1.2034 &   1.2838 &   1.3820
\\
0.36
 &   0.6947 &   1.2522 &   1.3405 &   1.4409
\\  \hline \hline
\end{tabular}}
\end{center}\label{tab3}
\end{table}
\begin{table}[h]
\caption{ String tension parameter $\sqrt{\sigma}$ (GeV)  on loop
numbers and normalization conditions. Normalization conditions
are the same as in Table~\ref{tab2}.}
\begin{center}
{\begin{tabular}{@{}ccccc@{}} \hline \hline $\alpha_s(m_{\tau}^2)$
&1-loop &2-loop&3-loop&4-loop
     \\ \hline \hline
0.29
 &   0.2098 &   0.3398 &   0.3297 &   0.3426
\\
0.30
 &   0.2274 &   0.3615 &   0.3522 &   0.3654
\\
0.31
 &   0.2451 &   0.3827 &   0.3745 &   0.3878
 \\
0.32
 &   0.2630 &   0.4035 &   0.3964 &   0.4099
\\
0.33
 &   0.2809 &   0.4239 &   0.4180 &   0.4315
\\
0.34
 &   0.2990 &   0.4437 &   0.4392 &   0.4527
\\
0.35
 &   0.3170 &   0.4629 &   0.4599 &   0.4733
\\
0.36
 &   0.3351 &   0.4817 &   0.4802 &   0.4935
\\ \hline \hline
\end{tabular}}
\end{center}\label{tab5}
\end{table}

\section{Gluon Condensate}
Let us turn to  calculation of the gluon condensate. Its value is
defined by the nonperturbative part of $\alpha_s$. We have (see,
e.g. the third of Refs.~\cite{AA})
\begin{equation}
K\equiv <\alpha_s/\pi: G^a_{\mu\nu} G^a_{\mu\nu}:>=
\frac{3}{\pi^3}\int\limits_{0}^{\infty} dq^2\,q^2 \alpha_{\rm
npt}(q^2). \label{111}
\end{equation}
We define the nonperturbative part of the invariant charge as a
difference of total invariant charge $\alpha_s$ and its
perturbative part (to all orders). In our approach the
nonperturbative contributions are  presented  for $q^2<q_0^2$
only. So we have
\begin{equation}
K=\frac{3}{\pi^3}\int\limits_{0}^{\infty} dq^2\,q^2(\alpha_s(q^2)-
\alpha_{\rm pt}(q^2)) =\frac{3}{\pi^3}\int\limits_{0}^{q_0^2}
dq^2\,q^2(\alpha_s(q^2)- \alpha_{\rm pt}(q^2)). \label{112}
\end{equation}
Let us consider two variants of the invariant charge perturbative
component behaviour.

\subsection{Freezing of perturbative component}

For the beginning we define the perturbative part in
nonperturbative region  basing on the assumption of freezing of
$\alpha_{\rm pt}$ at small $q^2$ (see, e.g. Ref.~\cite{Simonov}).
That is we assume
\begin{equation}
\alpha_{\rm pt}(q^2)\,=\, \alpha_s(q_0^2)\,=4\pi h_0, \,\,\,
q^2<q_0^2. \label{12}
\end{equation}
Using expressions~(\ref{9}), (\ref{12}), we have
$$
K_{\rm fr}=\frac{12}{\pi^2}\int\limits_{0}^{q_0^2} dq^2\,q^2\left(
\frac{C}{q^2}+z-h_0 \right)=\frac{12}{\pi^2}q_0^4\left(
\frac{C}{q_0^2}+\frac{1}{2}(z-h_0) \right)
$$
\begin{equation}
=\frac{6}{\pi^2}(h_0-z)x_0^2\Lambda^4_{QCD}.
\label{13}
\end{equation}
Expression~(\ref{13}) is valid for each of 1 --- 4-loop
approximations of $\alpha_s$ in  perturbative region and ratio
$K_{\rm fr}/\Lambda^4_{QCD}$ does not depend on  normalization
conditions for $\alpha_s$. The values of $K_{\rm
fr}^{1/4}/\Lambda_{QCD}$ calculated with the aid of
expression~(\ref{13}) are  presented in Table~\ref{tab6}.
\begin{table}[h]
\caption{{Dimensionless quantities $K_{\rm
fr}^{1/4}/\Lambda_{QCD}$, $y_0$, $\xi$,} $K_{\rm
an}^{1/4}/\Lambda_{QCD}$ on loop numbers, $n_f=$ 3.}
\begin{center}
{\begin{tabular}{@{}ccccc@{}} \hline \hline
  &1-loop &2-loop&3-loop&4-loop
     \\ \hline \hline
$K_{\rm fr}^{1/4}/\Lambda_{QCD}$
 &   0.9799 &   0.8977 &   0.9952 &   1.0709
\\
$y_0$
 &   1.0\hphantom{000} &   0.8299 &   2.2691 &   2.8710
\\
$\xi$
 &   7.3893 &   9.3702 &   4.5225 &   4.3297
\\
$K_{\rm an}^{1/4}/\Lambda_{QCD}$
 &   0.9373 &   0.8494 &   0.9353 &   1.0025
\\ \hline \hline
\end{tabular}}
\end{center}\label{tab6}
\end{table}
For normalized solutions values of the gluon condensate $K_{\rm
fr}^{1/4}$ are presented in Table~\ref{tab4}. It is seen that the
two --- four-loop results are quite stable with respect to loop
numbers and for  normalization condition $\alpha_s(m^2_{\tau})=$
0.33 one has  $K_{\rm fr}=$ (0.355 - 0.383 GeV)$^4$, which is
close to the conventional value~\cite{Shif} of the gluon
condensate (0.33 GeV)$^4$.

\begin{table}[h]
\caption{ Values of the gluon condensate $K_{\rm fr}^{1/4}$
(GeV)  on loop numbers and normalization conditions. Normalization
conditions are the same as in Table~\ref{tab2}.}
\begin{center}
{\begin{tabular}{@{}ccccc@{}} \hline \hline $\alpha_s(m_{\tau}^2)$
&1-loop &2-loop&3-loop&4-loop
     \\ \hline \hline
0.29
 &   0.1568 &   0.2844 &   0.2859 &   0.3038
\\
0.30
 &   0.1699 &   0.3025 &   0.3054 &   0.3240
\\
0.31
 &   0.1832 &   0.3203 &   0.3247 &   0.3439
 \\
0.32
 &   0.1965 &   0.3377 &   0.3437 &   0.3635
\\
0.33
 &   0.2099 &   0.3547 &   0.3624 &   0.3827
\\
0.34
 &   0.2234 &   0.3713 &   0.3808 &   0.4014
\\
0.35
 &   0.2369 &   0.3874 &   0.3988 &   0.4198
\\
0.36
 &   0.2504 &   0.4031 &   0.4164 &   0.4377
\\ \hline \hline
\end{tabular}}
\end{center}\label{tab4}
\end{table}

\subsection{Analytic  behavior of perturbative component}

Let us consider another variant of definition of the perturbative
part of $\alpha_s$  for $q^2<q^2_0$.  Namely instead of
freezing~(\ref{12}) we shall assume ``forced'' analytic behavior
of $\alpha_{\rm pt}$ in this region,
\begin{equation}
\alpha_{\rm pt}(q^2)\,=\, \alpha_{\rm an}(q^2), \;\;\; q^2<q_0^2.
\label{21}
\end{equation}
The main ideas of the analytic approach in  quantum field theory,
which allows one to overcome difficulties, connected with
nonphysical singularities in perturbative expressions, were
proposed in Refs.~\cite{Red,Bog}. The analytic approach is
successfully applied to QCD~\cite{SolShirTMF}. The forced analytic
running coupling constant is defined by the spectral
representation
\begin{equation}
a_{\rm an}(y)=\frac{1}{\pi}\int\limits_0^\infty
\frac{d\sigma}{y+\sigma}
\rho(\sigma),
\label{22}
\end{equation}
where spectral density $\rho(\sigma)= \Im a_{\rm an}(-\sigma-i0)=$
$\Im a(-\sigma-i0)$. For the perturbative solutions $a(x)$ of the
form~(\ref{16}) the two-loop analytic running coupling constant
and its nonperturbative part were studied in Ref.~\cite{PR}, the
three-loop case and the four-loop case where studied in
Refs.~\cite{YadFiz} and Refs.~\cite{A4}, respectively. Let us
write the spectral density up to the four-loop case.
\begin{equation}
\rho^{(1)}(\sigma)=
\frac{\pi}{t^2+\pi^2},
\label{23}
\end{equation}
\begin{equation}
\rho^{(2)}(\sigma)=\rho^{(1)}(\sigma)
-\frac{b}{(t^2+\pi^2)^2}\left[
2\pi t F_1(t)-\left(t^2-\pi^2\right)F_2(t)\right],
\label{24}
\end{equation}
$$
\rho^{(3)}(\sigma)=\rho^{(2)}(\sigma)+\frac{b^2}{(t^2+\pi^2)^3}
\left[\pi
\left(3t^2-\pi^2\right)\left(
F_1^2(t)-F_2^2(t)\right)\right.
$$
$$
-2t\left(t^2-3\pi^2\right)F_1(t)F_2(t)
-\pi\left(3t^2-\pi^2\right)F_1(t)
\left.+t\left(t^2-3\pi^2\right)F_2(t)\right.
$$
\begin{equation}
\left.
+\pi\kappa\left(3t^2-\pi^2\right)\right],
\label{25}
\end{equation}
$$
\rho^{(4)}(\sigma)=\rho^{(3)}(\sigma)-\frac{b^3}
{(t^2+\pi^2)^4}\left[\left(t^4-
6\pi^2t^2+\pi^4\right)\left(F_2^3(t)-3F_1^2(t)F_2(t)
\right)
\right.
$$
$$
+4\pi t\left(t^2-
\pi^2\right)\left(F_1^3(t)-3F_1(t)F_2^2(t)\right)
-10\pi t\left(t^2-\pi^2\right)\left(F_1^2(t)-F_2^2(t)
\right)
$$
$$
+5\left(t^4-
6\pi^2t^2+\pi^4\right)F_1(t)F_2(t)
+4\pi \left(1+3\kappa\right)t\left(
t^2-\pi^2\right)F_1(t)
$$
\begin{equation}
-\left.\left(1+3\kappa\right)
\left(t^4-6\pi^2t^2+\pi^4\right)F_2(t)+
4\pi \bar\kappa t\left(t^2-\pi^2\right)\right].
\label{26}
\end{equation}
Here $t=\ln(\sigma)$,
\begin{equation}
F_1(t)\equiv\frac{1}{2}\ln(t^2+\pi^2),\,\,\,
F_2(t)\equiv\arccos\frac{t}{\sqrt{t^2+\pi^2}}.
\label{27}
\end{equation}
Solving the equation\footnote{While solving this equation it is
convenient to use the method of Refs.~\cite{YadFiz}, in which
$\alpha_{\rm an}^{\rm npt}(y)$ is represented as a series in
inverse powers of $y$.}
\begin{equation}
a_{\rm an}(y)=b_0 h_0,
\label{28}
\end{equation}
we find values $y_0$ for $a_{\rm an}$, being defined by
formulas~(\ref{22}) -- (\ref{27}) with the use of values
$h_0$ obtained above.
Further we find dimensionless quantity  $\xi=$
$(\Lambda_{\rm an}/\Lambda_{QCD})^2=$ $x_0/y_0$. Values of
$y_0$ and $\xi$ in dependence on the number of loops are
presented in Table~\ref{tab6}.
In Table~\ref{tab7} values of $\Lambda_{\rm an}=$
$q_0/\sqrt{y_0}$ are presented in dependence on the
number  of loops and on normalization conditions at
$q^2=m_{\tau}^2$.

Let us turn to the gluon condensate. In the considered method of
definition of the perturbative part of $\alpha_s$ in the
nonperturbative region we have
$$
 K_{\rm an}=
\frac{3}{\pi^3}\int\limits_{0}^{q_0^2} dq^2\,q^2(\alpha_s(q^2)-
\alpha_{\rm an}(q^2))=\frac{12}{\pi^2}\int\limits_{0}^{q_0^2}
dq^2\,q^2\left( \frac{C}{q^2}+z\right.
$$
\begin{equation}
\left.-\frac{1}{b_0}a_{\rm an}\left(\frac{q^2} {\Lambda_{\rm
an}^2} \right) \right). \label{29}
\end{equation}
Taking into account the  spectral representation~(\ref{22})
and performing integration in Eq.~(\ref{29}), we obtain
\begin{equation}
K_{\rm an}=
\frac{12}{\pi^2}\left(C q_0^2+\frac{z}{2}q_0^4\right)
-\frac{12\Lambda^4_{\rm an}}{\pi^3 b_0}\int\limits_{0}^{\infty}
d\sigma\,\rho(\sigma)\left[y_0-\sigma\ln\left(1+\frac{y_0}
{\sigma}
\right)\right].
\label{30}
\end{equation}
Substitution $\sigma=\exp(t)$
leads to the following representation of the gluon condensate
$$
K_{\rm an}=
\Lambda_{QCD}^4\left[\frac{12}{\pi^2}\left(h_0-\frac{z}{2}\right)
x_0^2 -\frac{12\xi^2}{\pi^3 b_0}\int\limits_{-\infty}^{\infty} d
t\,\rho(t)\left\{y_0e^t \right.\right.
$$
\begin{equation}
\left.\left.
 -e^{2t}\ln\left(1+y_0e^{-t}\right)\right\} \right].
\label{31}
\end{equation}
The expression in the square brackets of~(\ref{31}) does
not depend on values of $\alpha_s(m_{\tau}^2)$, values of
the ratio $K_{\rm an}^{1/4}/\Lambda_{QCD}$, which are
obtained by numerical integration in formula~(\ref{31}),
are given in Table~\ref{tab6}.
In Table~\ref{tab7} and Table~\ref{tab8} values of the parameter
$\Lambda_{\rm an}$ and of the gluon condensate
$K_{\rm an}^{1/4}$ are presented, respectively.

\begin{table}[h]
\caption{ Dependence of the parameter $\Lambda_{\rm an}$ (GeV) on
loop numbers and normalization conditions. Normalization
conditions are the same as in
 Table~\ref{tab2}.}
\begin{center}
{\begin{tabular}{@{}ccccc@{}} \hline \hline
$\alpha_s(m_{\tau}^2)$  &1-loop &2-loop&3-loop&4-loop
     \\ \hline \hline
0.29
 &   0.4350 &   0.9696 &   0.6109 &   0.5903
\\
0.30
 &   0.4714 &   1.0315 &   0.6526 &   0.6296
\\
0.31
 &   0.5081 &   1.0921 &   0.6939 &   0.6682
\\
0.32
 &   0.5451 &   1.1515 &   0.7345 &   0.7063
\\
0.33
 &   0.5824 &   1.2094 &   0.7745 &   0.7435
\\
0.34
 &   0.6198 &   1.2659 &   0.8138 &   0.7800
\\
0.35
 &   0.6572 &   1.3210 &   0.8522 &   0.8156
\\
0.36
 &   0.6947 &   1.3745 &   0.8899 &   0.8504
\\ \hline \hline
\end{tabular}}
\end{center}\label{tab7}
\end{table}
\begin{table}[h]
\caption{ Values of the gluon condensate $K_{\rm an}^{1/4}$ (GeV)
in dependence on loop numbers and normalization conditions.
Normalization conditions are the same as in
 Table~\ref{tab2}.}
\begin{center}
{\begin{tabular}{@{}ccccc@{}} \hline \hline
$\alpha_s(m_{\tau}^2)$  &1-loop &2-loop&3-loop&4-loop
     \\ \hline \hline
0.29
 &   0.1500 &   0.2691 &   0.2687 &   0.2844
\\
0.30
 &   0.1625 &   0.2862 &   0.2870 &   0.3033
\\
0.31
 &   0.1752 &   0.3031 &   0.3052 &   0.3219
 \\
0.32
 &   0.1880 &   0.3195 &   0.3230 &   0.3403
\\
0.33
 &   0.2008 &   0.3356 &   0.3406 &   0.3582
\\
0.34
 &   0.2137 &   0.3513 &   0.3579 &   0.3758
\\
0.35
 &   0.2266 &   0.3666 &   0.3748 &   0.3929
\\
0.36
 &   0.2395 &   0.3814 &   0.3914 &   0.4097
\\ \hline \hline
\end{tabular}}
\end{center}\label{tab8}
\end{table}

\begin{figure}[th]
\centerline{\psfig{file=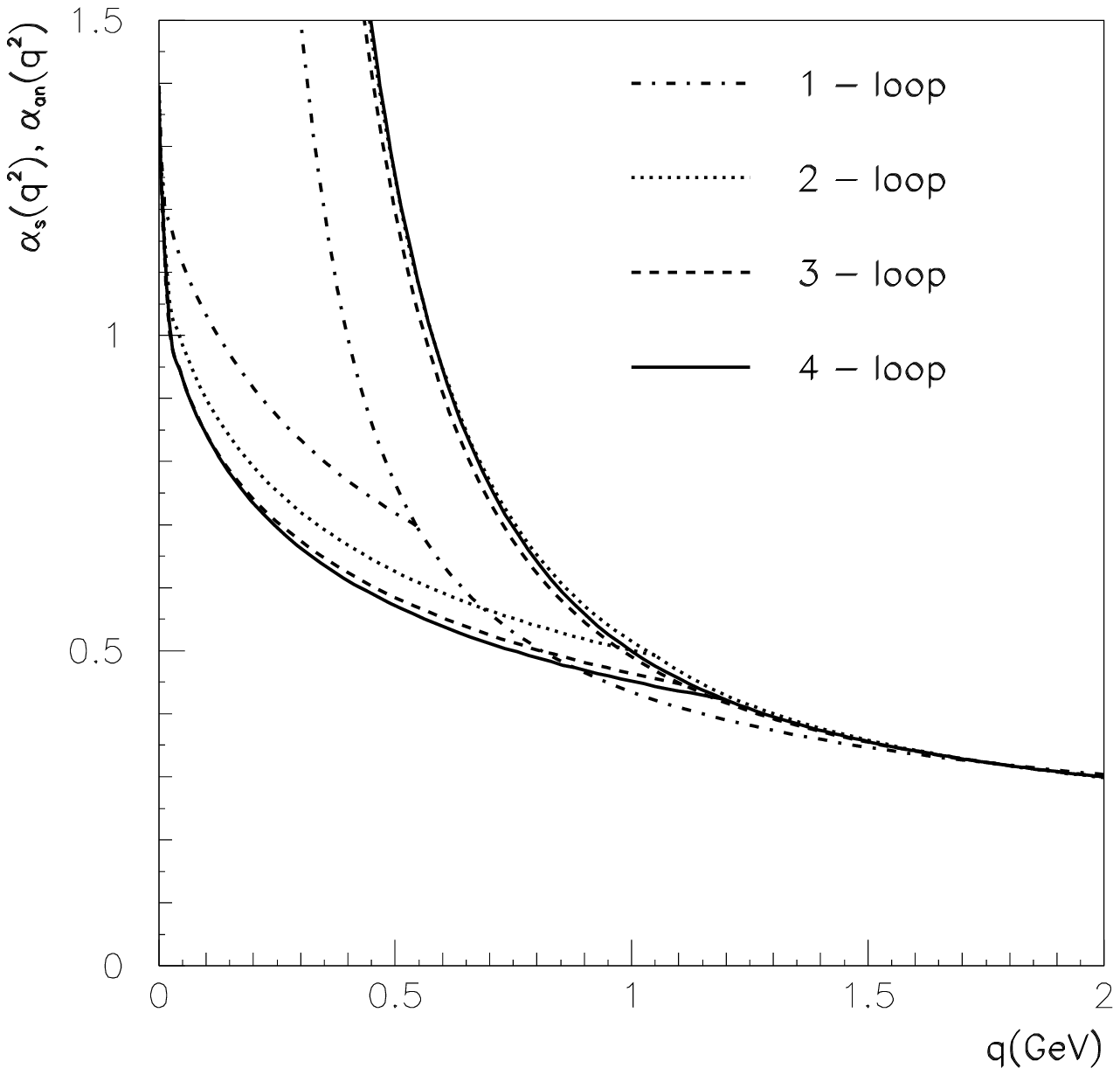,width=15cm}}
\vspace*{8pt}
\caption{Running coupling constant $\alpha_s(q^2)$ for
all $q^2>0$ and analytic coupling constant
$\alpha_{\rm an}(q^2)$ for $q^2<q^2_0$ (the corresponding curves
are the lower ones). Normalization conditions:
$\alpha_s(m_{\tau}^2)=$ 0.32, $\alpha_{\rm an}(q_0^2)=$ $\alpha_s
(q_0^2)$.}
\label{fig1}
\end{figure}

\begin{figure}[htb]
\centerline{\psfig{file=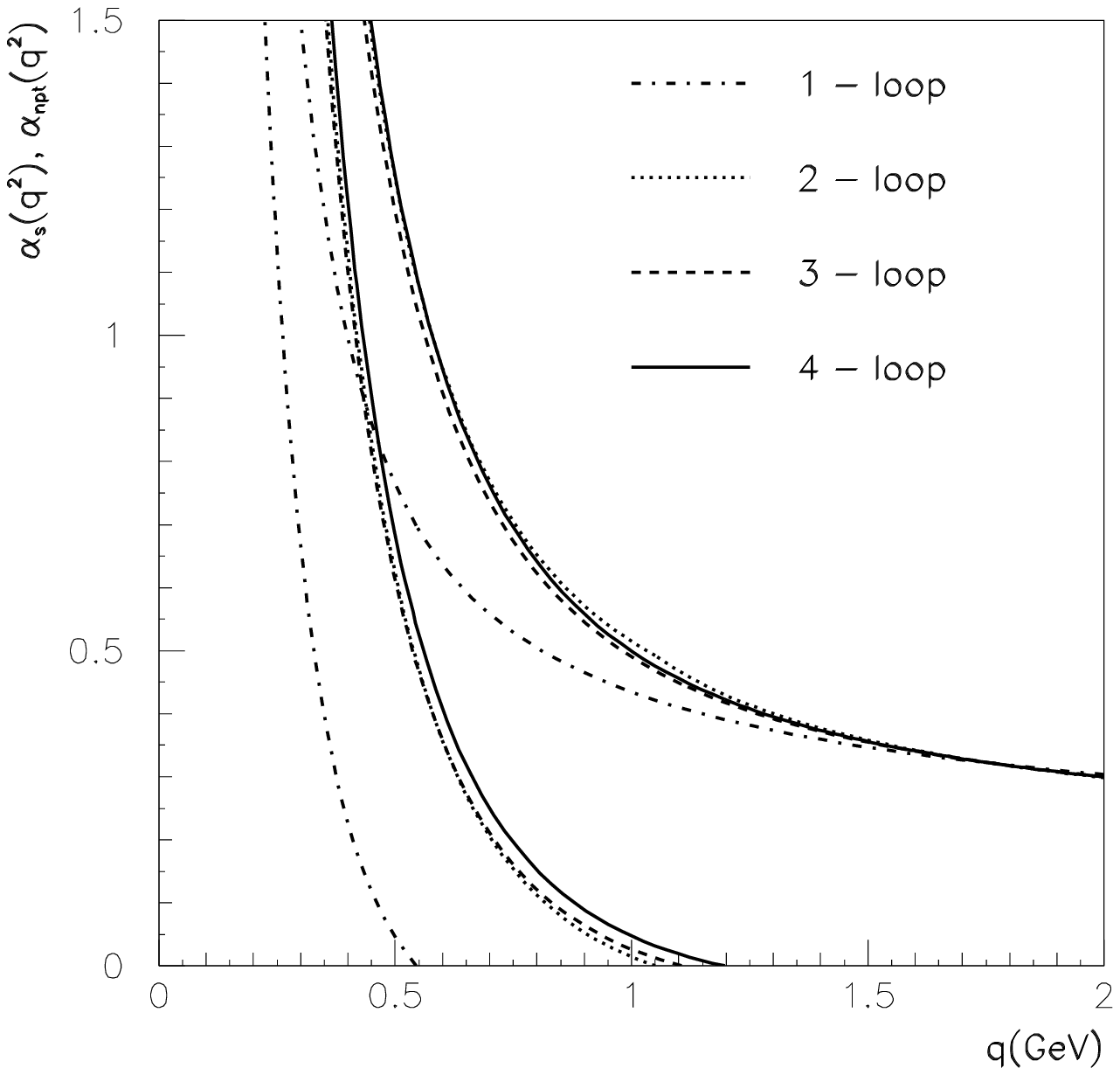,width=15cm}}
\vspace*{8pt}
\caption{Running coupling constant $\alpha_s(q^2)$ and
its nonperturbative part $\alpha_{\rm npt}(q^2)$ (the corresponding
curves are the lower ones) with definition of
$\alpha_{\rm pert}(q^2)$ by (\ref{21}). Normalization condition:
$\alpha_s(m_{\tau}^2) = 0.32$.}
\label{fig2}
\end{figure}

For cases from one loop up to four loops the behavior of the
running coupling constant $\alpha_s(q^2)$ for all $q^2>0$ is
shown in~Fig.~\ref{fig1}. Here the behavior of the analytic
coupling constant $\alpha_{\rm an}(q^2)$ for $q^2<q^2_0$, which
defines the perturbative part of $\alpha_s(q^2)$ in this
region, is also shown.
%
In~Fig.~\ref{fig2}
%
%
in addition to $\alpha_s(q^2)$ its nonperturbative part
$\alpha_{\rm npt}(q^2)$ is shown, which turns to be zero for $q^2
> q_0^2$. Normalization condition for all curves in~
Fig.~\ref{fig1},  Fig.~\ref{fig2} is $\alpha_s(m_\tau^2) = 0.32$.

\section{Conclusion}

Starting from the well-known perturbative expression (\ref{1}) for
the $\beta$-function for small values of the coupling constant
$h$ and the behavior (\ref{4}) for large values of the coupling
constant, which corresponds to the linear confinement, we have
constructed the model $\beta$-function for all $h > 0$. To
control dependence of the results on the number of loops we
simultaneously consider cases corresponding to perturbative
$\beta$-function for 1 --- 4 loops.

While constructing the matched $\beta$-function we assume it to
be smooth at the matching point, that leads to the invariant
charge being smooth together with its derivative for all $q^2 >
0$. The normalization of the invariant charge, e.g. at  $m_\tau$
fix it thoroughly. Then value   $\alpha_s(m_\tau^2) \simeq 0.33$
corresponds to value of the parameter $\sigma^{1/2} \simeq 0.42$
GeV of our model which fits the string tension parameter of the
string  model~\cite{Sol}.

The obtained invariant charge is applied to  study of the
important physical quantity, the gluon condensate. In doing this
we consider two variants of extracting of the nonperturbative
contributions from the overall expression for the invariant
charge. The first variant assumes ``freezing" of perturbative
part of the charge in the nonperturbative region $q^2 <
q_0^2$,\footnote{ It is seen from Table~1 that  freezing occurs
at the level $\alpha_s =\alpha_0 \simeq 0.4354$ (average for
3-loop and 4-loop cases) which is very close to critical value
$\alpha_{\rm crit} =0.137\cdot \pi=0.4304$ of Ref.~\cite{Doksh}.}
while for the second one we choose the analytic behavior of the
perturbative part in this region. As we see from Tables~
\ref{tab4}, \ref{tab8} the first variant leads to values of the
gluon condensate being somewhat larger, than that for the second
variant.

Emphasize, that for $\alpha_s(m_\tau^2) = 0.33$ values of the
gluon condensate for both  variants are quite satisfactory.
Namely the value for the second variant practically coincides the
conventional value~\cite{Shif}, while for the first variant it is
only slightly higher,  $K^{1/4}=$ $K_{\rm fr}^{1/4} \simeq 0.36 $
GeV. Note, that other important parameter, the nonperturbative
scale $q_0$ for the same normalization condition also turns to be
of a reasonable magnitude, $q_0 \simeq 1.17$ GeV (see
Table~\ref{tab3}). We may conclude, that the present model
consistently describes the most important nonperturbative
parameters. From this point of view the results support a
possibility of singular infrared behaviour~(\ref{3}) of the
invariant charge which contains both perturbative and
nonperturbative contributions.

\section*{Acknowledgments}

The authors express deep gratitude to V.A. Petrov, V.E. Rochev
and D.V. Shirkov for valuable discussions. Moreover, we thank A.L.
Kataev and A.A. Pivovarov for stimulating discussions and
comments. The work has been partially supported by RFBR under
Grant No. 02-01-00601.

\end{document}